# High Thermoelectric Performance Originating from the Grooved Bands in the ZrSe$_3$ Monolayer

Zizhen Zhou, Huijun Liu[*], Dengdong Fan, Guohua Cao, and Caiyu Sheng

*Key Laboratory of Artificial Micro- and Nano-Structures of Ministry of Education and School of Physics and Technology, Wuhan University, Wuhan 430072, China*

Low-dimensional layered materials have attracted tremendous attentions due to their wide range of physical and chemical properties and potential applications in electronic devices. Using first-principles method taking into account the quasiparticle self-energy correction and Boltzmann transport theory, the electronic transport properties of ZrSe$_3$ monolayer are investigated, where the carrier relaxation time is accurately calculated within the framework of electron-phonon coupling. It is demonstrated that the high power factor of the monolayer can be attributed to the grooved bands near the conduction band minimum. Combined with the low lattice thermal conductivity obtained by solving the phonon Boltzmann transport equation, a considerable *n*-type *ZT* value of ~2.4 can be achieved at 800 K in the ZrSe$_3$ monolayer.

## 1. Introduction

With the world's increasing demand for energy, a compelling need exists for high performance thermoelectric materials which can directly convert waste heat into electrical power. The thermoelectric conversion efficiency is usually determined by the figure-of-merit $ZT = S^2\sigma T/\kappa$, where $S$ is the Seebeck coefficient, $\sigma$ is the electrical conductivity, $T$ is the absolute temperature, and $\kappa$ is the sum of the electronic ($\kappa_e$) and lattice ($\kappa_l$) thermal conductivity. To achieve a high *ZT* value, a thermoelectric material requires large power factor ($S^2\sigma$) and/or low thermal conductivity. During the past two decades, various effective strategies have been applied to improve the efficiency of thermoelectric materials [1−3]. In particular, the pioneering work of Hicks *et al.* [4, 5] proposed that low-dimensional structures could realize high thermoelectric performance by maintaining high power factor and low thermal conductivity at the same time.

---

[*] Author to whom correspondence should be addressed. Electronic mail: phlhj@whu.edu.cn

Among various low-dimensional systems, two-dimensional layered materials can be prepared via liquid phase or mechanical exfoliation of their bulk structures which are typically consist of stacked layers [6, 7] bonded together with weak van der Waals (vdW) interactions. Transition metal trichalcogenides (TMTCs) $MX_3$ are typical vdW stacked layered materials belonging to space group $P2_1/m$, where $M$ is the transition metal elements Ti, Zr or Hf and $X$ is S, Se or Te [8]. It is thus natural to ask whether the two-dimensional $MX_3$ monolayers could be obtained by exfoliating their bulk counterparts. Indeed, Jin *et al.* [9] theoretically suggested that the single layers of $TiS_3$, $TiSe_3$, $ZrS_3$, and $ZrSe_3$ could be exfoliated from their bulk crystals due to the low cleavage energies. Experimentally, the $TiS_3$ monolayer has been successfully isolated and exhibits a direct band gap of 1.1 eV [10−14]. Besides, Osada *et al.* [15] prepared the few-layers $ZrS_3$ and $ZrSe_3$, and their phonon properties were characterized by Raman spectroscopy. Recently, several first-principles approaches have been carried out to calculate the electronic transport coefficients of $MX_3$ monolayers, which indicate that they may be potential thermoelectric materials [9, 16]. It was suggested by Dai *et al.* [17] that nearly all of the $MX_3$ monolayers are semiconducting with band gaps in the range of 0.2~2.0 eV, while the only exception is the metallic $MTe_3$. Furthermore, Zhang *et al.* [18] demonstrated that the $TiS_3$ monolayer exhibits very high thermoelectric performance, which is superior to that of the bulk system. It is thus interesting to check if other kinds of $MX_3$ monolayers could also be potential thermoelectric materials, and a comprehensive understanding is quite necessary.

In this study, the electronic, phonon, and thermoelectric transport properties of $ZrSe_3$ monolayer are investigated by using first-principles calculations and Boltzmann transport theory. We shall see that the thermoelectric performance of the system exhibits strong anisotropy, and an *n*-type $ZT$ value as high as 2.4 could be achieved at 800 K at moderate carrier concentration. Such superior thermoelectric performance can be traced back to the grooved bands near the conduction band minimum.

## 2. Methodology

The electronic properties of ZrSe$_3$ monolayer have been investigated within the framework of density functional theory (DFT) [19, 20], which is implemented in the so-called Vienna *Ab-initio* Simulation Package (VASP) code [21]. The exchange-correlation functional is in the form of Perdew-Burke-Ernzerhof (PBE) with generalized gradient approximation (GGA) [22]. In addition, the *GW* approximation is considered to accurately predict the band structure [23−26]. The electronic transport coefficients are computed by using the Boltzmann transport theory [27] with the maximally localized Wannier function [28−30] basis to interpolate the *GW* band structure [31], where a very dense *k*-mesh of 9216 points is used in the whole Brillouin zone. Here the key point is appropriate treatment of the carrier relaxation time since complex scattering mechanisms are usually involved. Earlier attempts for this issue adopted the deformation potential theory [32], where the relaxation time is treated as a constant [33, 34] and the calculated value is generally overestimated [35, 36]. In the present work, we obtain the relaxation time from the imaginary part of the electron self-energy by a complete electron-phonon coupling (EPC) [37] calculation, as done in the electron-phonon Wannier (EPW) [38] package. In this approach, the *k*-resolved relaxation time is given by

$$\frac{1}{\tau_{n\mathbf{k}}(\mu,T)} = \frac{2\pi}{\hbar} \sum_{m,\mathbf{q}v} \left|g_{mn}^{v}(\mathbf{k},\mathbf{q})\right|^{2} \{[n(\omega_{\mathbf{q}v},T) + f(\varepsilon_{m\mathbf{k+q}},\mu,T)]\delta(\varepsilon_{n\mathbf{k}} - \varepsilon_{m\mathbf{k+q}} + \hbar\omega_{\mathbf{q}v}) \\ + [n(\omega_{\mathbf{q}v},T) + 1 - f(\varepsilon_{m\mathbf{k+q}},\mu,T)]\delta(\varepsilon_{n\mathbf{k}} - \varepsilon_{m\mathbf{k+q}} - \hbar\omega_{\mathbf{q}v})\}, \quad (1)$$

where $\varepsilon_{m\mathbf{k}}$ is the energy eigenvalue of band *m* and wavevector *k*, and $\omega_{\mathbf{q}v}$ is the frequency of a phonon mode at wavevector *q* and polarization *v*. $f(\varepsilon_{m\mathbf{k+q}},\mu,T)$ and $n(\omega_{\mathbf{q}v},T)$ are respectively the Fermi-Dirac and Bose-Einstein distribution function, in which the temperature (*T*) and chemical potential (*μ*) dependence are included. The EPC matrix element is defined as $g_{mn}^{v}(\mathbf{k},\mathbf{q}) = \langle \psi_{m\mathbf{k+q}} | \partial_{\mathbf{q}v} V | \psi_{n\mathbf{k}} \rangle$, where $\psi_{m\mathbf{k}}$ and $\partial_{\mathbf{q}v} V$ are the wavefunction and self-consistent potential, respectively. In order to achieve converged carrier relaxation time, the EPC calculations have been performed by using coarse grids of $8\times8\times1$ ***k***-points and $4\times4\times1$ ***q***-points, and then interpolated with dense meshes of $96\times96\times1$ ***k***-points and $48\times48\times1$ ***q***-points.

For the phonon transport calculations, the harmonic and anharmonic properties are investigated by using the finite displacement method as implemented in the Phonopy package [39] and the Thirdorder.py program, respectively. A $4\times4\times1$ supercell with *Γ* point is employed to calculate the second- and third-order interatomic force constants (IFCs). Besides, the interactions are set to be the 8$^{th}$ nearest neighbors to obtain convergent results for the anharmonic IFCs. The lattice thermal conductivity can be obtained by solving the phonon Boltzmann transport equation implemented in the so-called ShengBTE code [40], and a fine $96\times96\times1$ ***q***-mesh is adopted to ensure convergence.

## 3. Results and discussion

As discussed above, the monolayer ZrSe$_3$ is expected to be obtained by mechanical cleavage or liquid phase exfoliation of its bulk counterpart. The top and side views of the monolayer are shown in Figure 1(a) and 1(b), respectively. The optimized lattice constants of the primitive cell are respectively 5.48 Å and 3.78 Å along the *x*- and *y*-direction, and the thickness is as large as 6.32 Å. Our calculated lattice parameters are in good agreement with those reported previously [9]. As can be seen from Fig. 1(b), there are two types of Se atoms in the system, where the outer and inner ones are marked as Se1 and Se2, respectively. The outer Se atoms form Se1-Se1 chain with bond length of $d_1$ = 2.38 Å, and Zr-Se1 bond with length of $d_2$ = 2.77 Å. In addition, there are two kinds of Zr-Se2 bonds along and across the plane with lengths of $d_3$ = 2.91 Å and $d_4$ = 2.75 Å, respectively. Such mixed covalent bonds in the ZrSe$_3$ monolayer suggest that it may exhibit relatively lower lattice thermal conductivity, which is beneficial to high thermoelectric performance.

Figure 2(a) plots the energy band structures of ZrSe$_3$ monolayer, where the results calculated with the PBE functional and *GW* approximation are both shown for comparison. The PBE bands show an indirect gap of 0.39 eV based on a careful search in the whole Brillouin zone, where the conduction band minimum (CBM) and valence band maximum (VBM) are located at the **k** points of (0.500, 0.177, 0.000) and (0.000, 0.000, 0.000), respectively. It is well known that the standard DFT tends to underestimate the band gap seriously, and such limitation can be solved by calculating the quasiparticle properties with the *GW* approximation of the many-body effects. As can be found from Fig. 2(a), the major change caused by *GW* calculation is that the conduction band is obviously upshifted and the energy gap is significantly increased to 1.63 eV compared with the PBE result. If we focus on the conduction band bottom, we see that the band dispersion along the *YS* direction (*x*-direction) is much larger than that along the *SX* direction (*y*-direction). In another word, there is a mixture of light and heavy conduction bands which is very beneficial for achieving high thermoelectric performance [41]. To have a better understanding, the energy dispersion relations of the top valence band and bottom conduction band are shown in Fig. 2(b). It is clear that the bottom conduction band is rather flat along the *y*-direction but quite steep along the *x*-direction. Such unique groove-like band structure could have important influence on the electronic transport properties of ZrSe$_3$ monolayer, as will be discussed in the following.

Within the framework of Boltzmann transport theory, the Seebeck coefficient, the electrical conductivity, and the electronic thermal conductivity of ZrSe$_3$ monolayer can be expressed as

$$S(\mu,T) = -\frac{1}{eT} \frac{\sum_{n\mathbf{k}} (\varepsilon_{n\mathbf{k}} - \mu) \mathbf{v}_{n\mathbf{k}}^2 \tau_{n\mathbf{k}}(\mu,T) \frac{\partial f(\varepsilon_{n\mathbf{k}},\mu,T)}{\partial \varepsilon}}{\sum_{n\mathbf{k}} \mathbf{v}_{n\mathbf{k}}^2 \tau_{n\mathbf{k}}(\mu,T) \frac{\partial f(\varepsilon_{n\mathbf{k}},\mu,T)}{\partial \varepsilon}}, \quad (2)$$

$$\sigma(\mu,T) = \frac{1}{N_\mathbf{k} V} \sum_{n\mathbf{k}} -e^2 \mathbf{v}_{n\mathbf{k}}^2 \tau_{n\mathbf{k}}(\mu,T) \frac{\partial f(\varepsilon_{n\mathbf{k}},\mu,T)}{\partial \varepsilon}, \quad (3)$$

$$\kappa_e(\mu,T) = \frac{1}{N_\mathbf{k} V} \sum_{n,\mathbf{k}} -\frac{(\varepsilon_{n\mathbf{k}} - \mu)^2}{T} \mathbf{v}_{n\mathbf{k}}^2 \tau_{n\mathbf{k}}(\mu,T) \frac{\partial f(\varepsilon_{n\mathbf{k}},\mu,T)}{\partial \varepsilon} - TS^2(\mu,T)\sigma(\mu,T), \quad (4)$$

Here *e* is the electron charge, *T* is the absolute temperature, $\mu$ is the chemical potential (corresponds to the carrier concentration), $N_\mathbf{k}$ is the total number of

***k***-points, $V$ is the volume of the primitive cell, $\mathbf{v}_{n\mathbf{k}}$ and $\varepsilon_{n\mathbf{k}}$ are respectively the group velocity and eigenvalue with band index $n$ at state ***k***, and $f(\varepsilon_{n\mathbf{k}}, \mu, T)$ is the Fermi-Dirac distribution function. As discussed above, the electron relaxation time $\tau_{n\mathbf{k}}(\mu, T)$ is accurately predicted by a complete EPC calculation. It should be noted that the transport coefficients depend on the definition of layer thickness, which is assumed to be the same as the interlayer separation of bulk ZrSe$_3$ (9.38 Å).

Figure 3(a) plots the energy dependence of the carrier relaxation time at two typical temperatures of 300 and 800 K. We do not consider higher temperature since the melting point of ZrSe$_3$ single layer is estimated to be lower than 810 K as derived from the decomposition temperature of the TiS$_3$ monolayer with the same crystal structure [17]. We find that the relaxation time at 300 K is obviously larger than that at 800 K, and they follow almost the same energy dependence. Besides, it can be seen that the relaxation time is relatively larger around the band edges, where the scattering channels are strongly limited due to the lower density of states (DOS). Within a narrow energy window around the conduction band edge, we detect a dramatic decrease of the electron relaxation time, which can be attributed to the sharp increase of the DOS caused by the heavy conduction band along the *SX* direction (see Fig. 2(a)). According to the Mott relation, the Seebeck coefficient can be written as

$$S = \frac{\pi^2 k_B^2 T}{3q} \left\{ \frac{1}{n} \frac{dn(\varepsilon)}{d\varepsilon} + \frac{1}{\mu} \frac{d\mu(\varepsilon)}{d\varepsilon} \right\}. \qquad (5)$$

where $n(\varepsilon)$ and $\mu(\varepsilon)$ are the carrier density (the product of DOS and $f(\varepsilon_{n\mathbf{k}}, \mu, T)$) and mobility, respectively [42]. The strong energy dependence of the DOS discussed above can thus lead to a large Seebeck coefficient, as also demonstrated previously [43]. On the other hand, the electrical conductivity of ZrSe$_3$ monolayer exhibits strong anisotropy due to significant difference of the band dispersion in the *x*- and *y*-direction. A much higher electrical conductivity can be obtained in the *x*-direction caused by larger group velocity. As a result, a considerably large power factor can be achieved along the *x*-direction due to simultaneously large

Seebeck coefficient and electrical conductivity. Indeed, our first-principles calculations find that the room temperature power factor in the *x*-direction is 50 times as much as that in the *y*-direction at optimized electron concentration. Fig. 3(b)~(d) show the electronic transport coefficients of *n*-type ZrSe$_3$ monolayer along the *x*-direction, where the results at 300 and 800 K are both plotted as a function of carrier concentration. At moderate electron concentration ($\sim 3\times 10^{12}$ cm$^{-2}$), we see from Fig. 3(b) that the absolute values of the Seebeck coefficients exceed 250 μV/K for both 300 and 800 K, as previously found in many good thermoelectric materials [44, 45]. Fig. 3(c) displays the electrical conductivity (solid lines) and the electronic thermal conductivity (dashed lines) of ZrSe$_3$ monolayer, where we find they follow almost the same carrier concentration dependence at 300 K as governed by the Wiedemann-Franz law [46]. This is however not the case at 800 K, since the last term $TS^2\sigma$ in Eq. (4) becomes important at high temperature. Due to the competitive behavior of the Seebeck coefficient and the electrical conductivity, a compromise must be taken to maximize the power factor, as shown in Fig. 3(d). Compared with that at 800 K, the maximum power factor at 300 K appears at a lower electron concentration, since more states in the conduction band contribute to the carrier concentration at higher temperature [47]. At the optimized electron concentration of $3.8\times 10^{12}$ cm$^{-2}$ ($2.9\times 10^{13}$ cm$^{-2}$), the power factor of ZrSe$_3$ monolayer can reach $8.2\times 10^{-3}$ W/mK$^2$ ($5.3\times 10^{-3}$ W/mK$^2$) at 300 K (800 K), which are higher than those of many good thermoelectric material such as Bi$_2$Te$_3$ [44] and SnSe [45] and convincingly confirms that the grooved bands can lead to better thermoelectric performance.

We now move to the discussion of the phonon transport properties of the ZrSe$_3$ monolayer. In the phonon dispersion relations shown in Figure 4(a), we see there are several low-frequency optic branches mixed with the acoustic ones, which is usually found in many systems with intrinsically low thermal conductivity [48]. Such a hybrid characteristic of phonon bands in the frequency range of 0~100 cm$^{-1}$ are mainly contributed by the surface Se1 atoms, as indicated in the corresponding phonon density of states (PDOS). Besides, it can be found that the dispersion of acoustic phonon branch along the $\Gamma Y$ direction is relatively stronger than that along the $\Gamma X$ direction, which suggests a smaller phonon group velocity and thus lower lattice thermal conductivity along the $x$-direction. Indeed, we see from Fig. 4(b) that the lattice thermal conductivity $\kappa_L$ of the ZrSe$_3$ monolayer exhibit obvious direction dependence, and the values along the $y$-direction are almost two times larger than those along the $x$-direction within the temperature range of 300~800 K. At room temperature, the calculated $\kappa_L$ are 2.9 and 8.3 W/mK along the $x$- and $y$-direction, respectively, which are comparable to those of the Bi$_2$Te$_3$ quintuple layer [49, 50] and suggest the favorable thermoelectric performance of ZrSe$_3$ monolayer. In addition, we have calculated the accumulative lattice thermal conductivity as a function of phonon frequency at 300 K. As shown in the inset of Fig. 4(b), the heat transport along the $x$- and $y$-direction are mainly contributed by the phonons in the frequency region of 0~100 cm$^{-1}$, where the corresponding PDOS is dominated by the outer Se1 atoms as discussed above. Namely, the Se1-Se1 bonds play a very important role in governing the heat transport of the ZrSe$_3$ monolayer. To have a better understanding, we plot in Fig. 4(c) the room temperature anharmonic three-phonon scattering rates (ASRs) versus phonon frequency. Compared with those of the phonon modes in the range of 0~100 cm$^{-1}$, the ASRs are much larger for the higher frequency phonons where the scattering is mainly caused by the Zr-Se1 and Zr-Se2 bonds. As known, the ASRs are determined by the anharmonic IFCs and the weighted phase space, where the former ones are usually characterized by the Gruneisen parameter. Detailed analysis reveals that the phase space of ZrSe$_3$ monolayer shows weak frequency dependence. In contrast, the Gruneisen parameters of the higher frequency phonons are obviously bigger than those of phonons with frequency of 0~100 cm$^{-1}$, which is consistent with the larger length of Zr-Se1 and Zr-Se2 bonds as compared with that of Se1-Se1 bond. Due to the same reason, we see in Fig. 4(d) that the phonon group velocities are lower for the high frequency branches than those in the range of 0~100 cm$^{-1}$. It should be mentioned that the EPC may also have certain effects on the phonon transport properties [51−54], especially at high carrier concentration. However, our additional calculations find that the phonon relaxation time originating from the EPC is at least two orders of magnitude larger than that from the intrinsic phonon-phonon scattering. It is thus reasonable to ignore the effects of EPC on the lattice thermal conductivity of ZrSe$_3$ monolayer.

With all the electronic and phonon transport coefficients obtained, we can now predict the thermoelectric performance of ZrSe$_3$ monolayer. Figure 5(a) shows the *n*-type *ZT* values along the *x*-direction, plotted as a function of carrier concentration at both 300 and 800 K. The corresponding transport coefficients are summarized in Table 1. We find that at the optimized electron concentration of $2.9\times10^{13}$ cm$^{-2}$ ($3.8\times10^{12}$ cm$^{-2}$), a maximum *ZT* value of ~2.4 (~0.7) can be obtained at 800 K (300 K). As can be seen from Fig. 5(b), the optimized *ZT* value of the *n*-type system in the *x*-direction increases almost linearly with the temperature, which suggests that enhanced thermoelectric performance may be realized in the ZrSe3 monolayer if it is operated in the high temperature region.

**4. Summary**

In summary, we present a comprehensive theoretical study on the thermoelectric properties of ZrSe$_3$ monolayer within the framework of DFT. It is found that an indirect band gap of 1.63 eV can be obtained by considering the quasiparticle self-energy correction. Detailed analysis of the band structure in the whole Brillouin zone reveals that the grooved conduction bands can lead to strong anisotropy of the electronic transport properties of the *n*-type system. In particular, a high Seebeck coefficient and electrical conductivity can be simultaneously achieved in the *x*-direction. Moreover, it is found that the heat transport in the ZrSe$_3$ monolayer is almost entirely contributed by the surface Se atoms. The relatively large bond length of Zr-Se1 and Zr-Se2 chains greatly limit their contribution to the heat transport, which directly leads to the low lattice thermal conductivity of the system. As a result, a maximum *ZT* value of ~2.4 can be realized at 800 K along the *x*-direction, suggesting the tremendous advantages of utilization of groove-like band structure in potential thermoelectric materials.

**Acknowledgements**

We thank financial support from the National Natural Science Foundation (Grant Nos. 11574236 and 51772220). The numerical calculations in this work have been done on the platform in the Supercomputing Center of Wuhan University.

**Table 1** The optimized *ZT* values of *n*-type ZrSe$_3$ monolayer along the *x*-direction at 300 and 800 K. The corresponding carrier concentration and transport coefficients are also listed.

| T (K) | $n$ ($10^{12}$ cm$^{-2}$) | $S$ (μV/K) | $\sigma$ (S/cm) | $S^2\sigma$ ($10^{-3}$ W/mK$^2$) | $\kappa_e$ (W/mK) | $\kappa_L$ (W/mK) | ZT |
|---|---|---|---|---|---|---|---|
| 300 | 3.8 | 226 | 1611 | 8.2 | 0.72 | 2.9 | 0.7 |
| 800 | 29 | 283 | 666.2 | 5.3 | 0.69 | 1.1 | 2.4 |

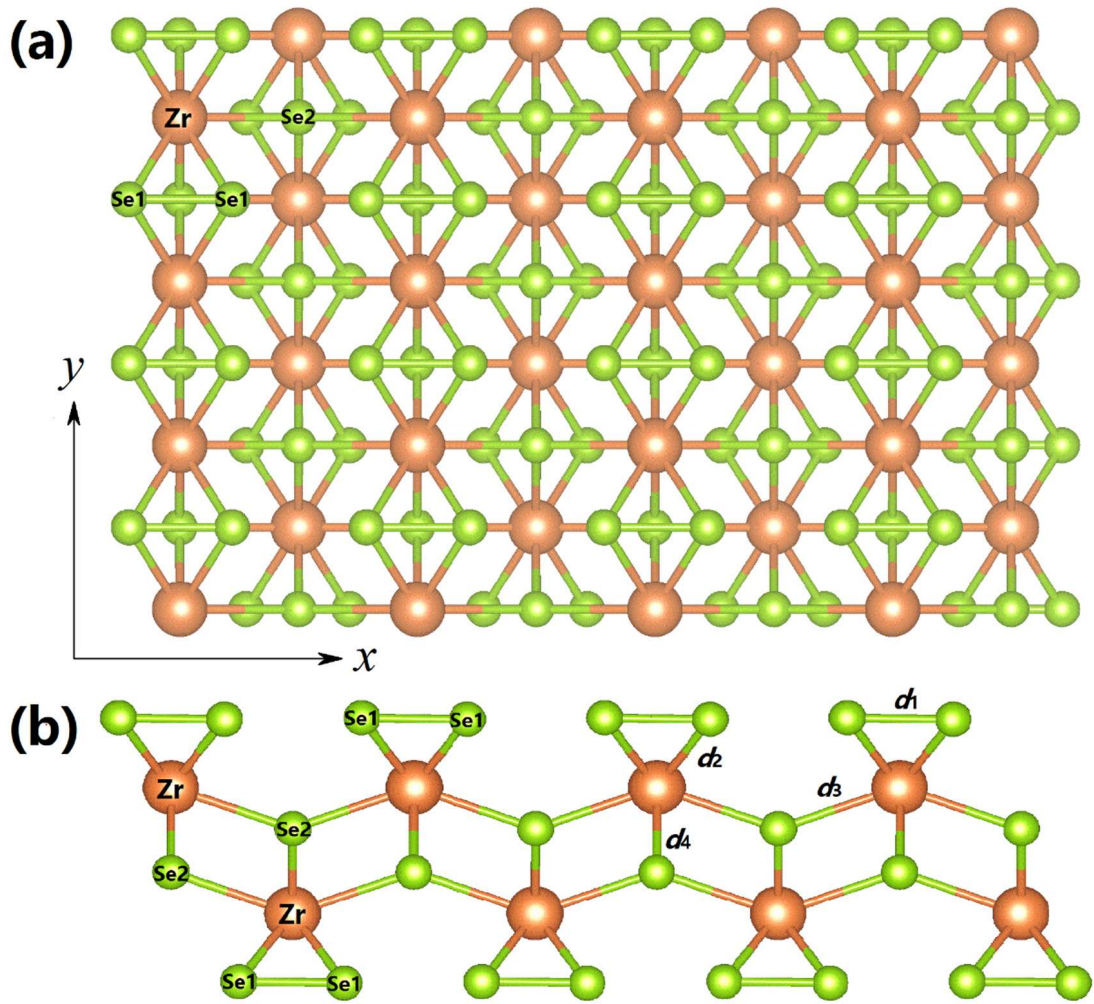

**Figure 1.** Ball-and-stick model of ZrSe₃ monolayer: (a) top-view and (b) side-view.

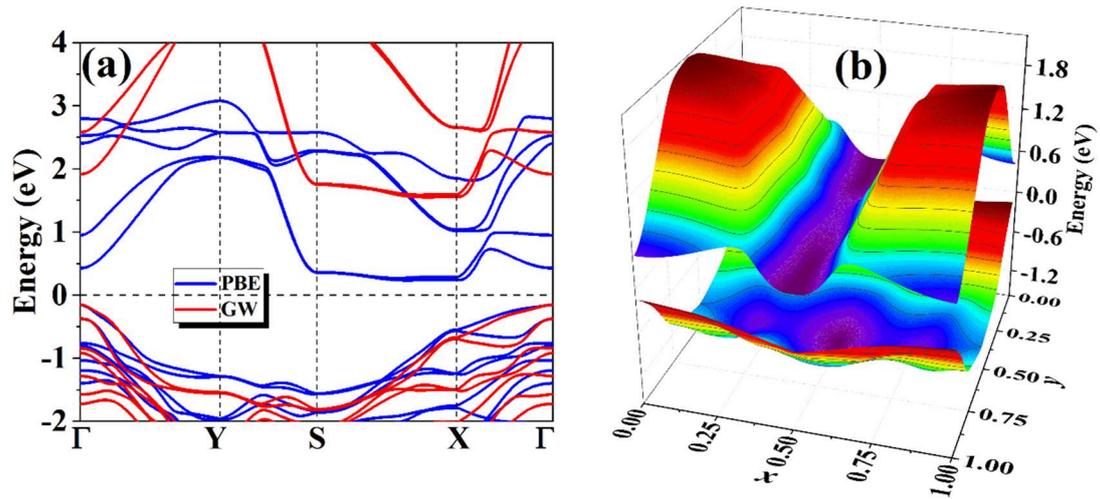

**Figure 2.** (a) The band structures of ZrSe$_3$ monolayer calculated with the PBE functional (blue lines) and *GW* approximation (red lines). (b) The three-dimensional energy dispersion relations of the top valence band and bottom conduction band.

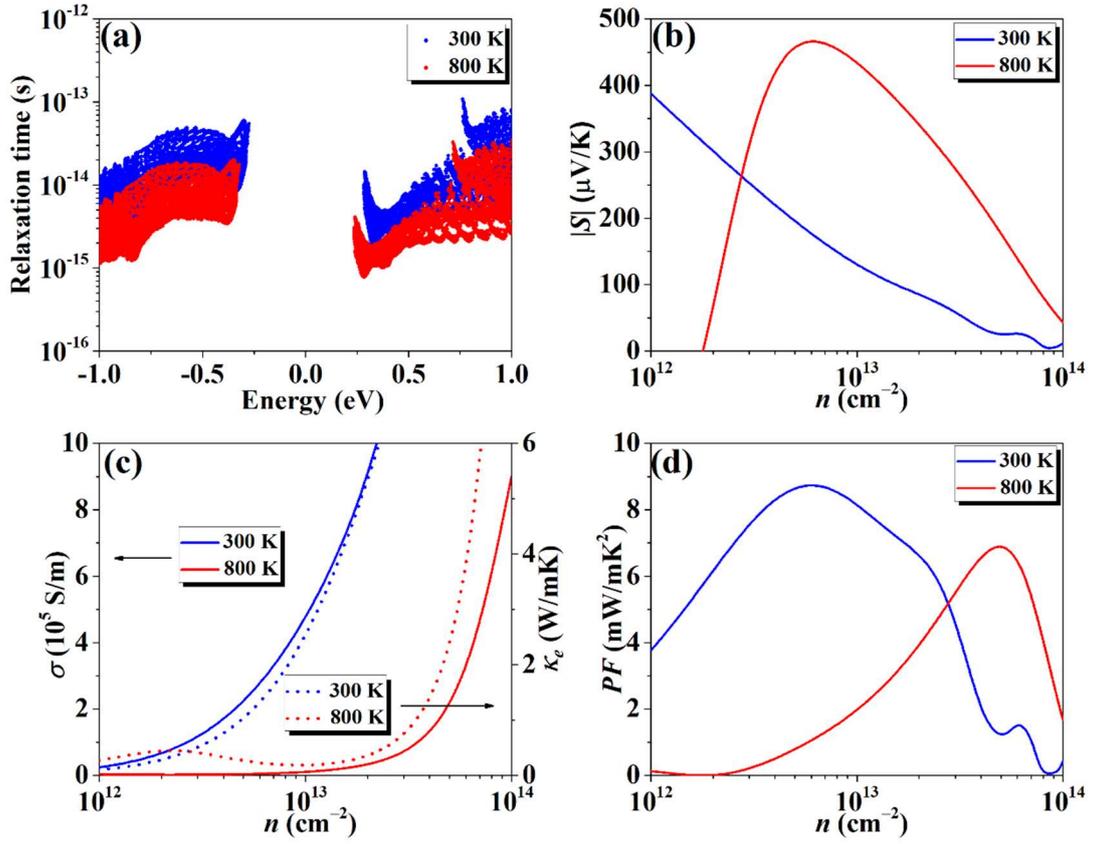

**Figure 3.** (a) The energy-dependent carrier relaxation time of ZrSe$_3$ monolayer at 300 and 800 K. The Fermi level is at 0 eV. (b) The absolute values of the Seebeck coefficient, (c) the electrical conductivity and electronic thermal conductivity, and (d) the power factor, all plotted as a function of electron concentration at 300 and 800 K along the *x*-direction.

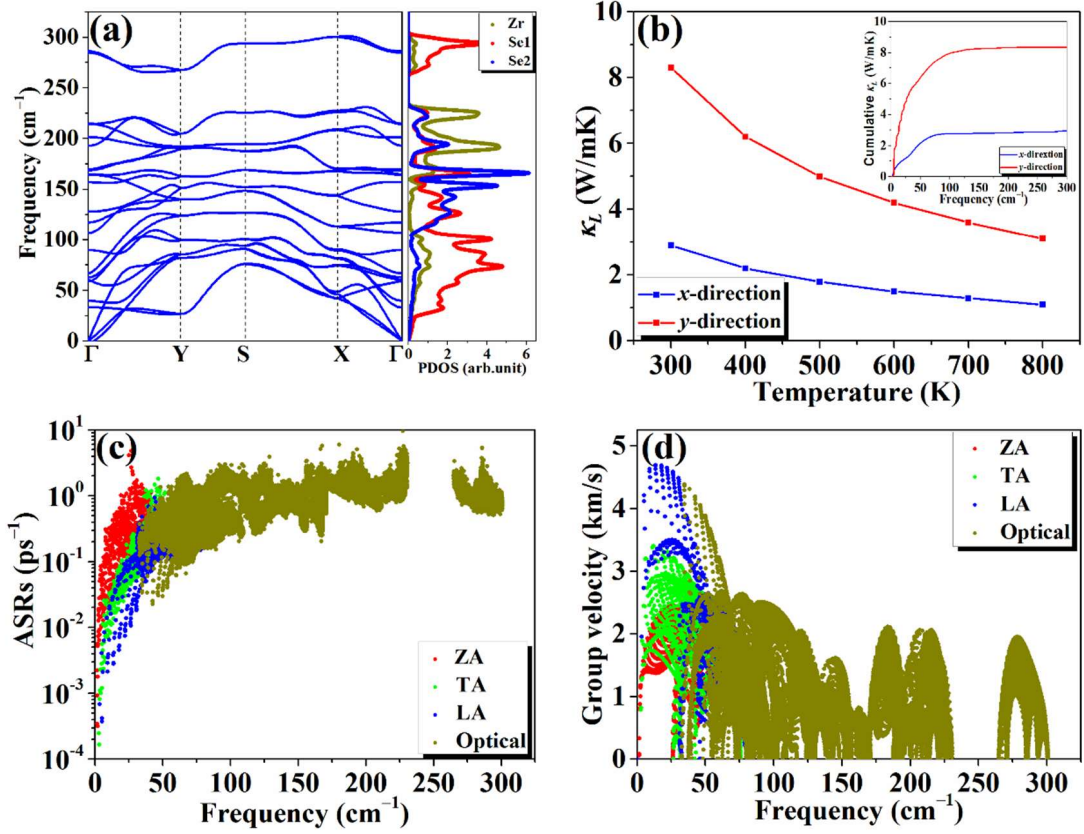

**Figure 4.** (a) The phonon dispersion relations and PDOS of ZrSe$_3$ monolayer. (b) The lattice thermal conductivity as a function of temperature. The inset shows the accumulative lattice thermal conductivity with respect to phonon frequency at 300 K. (c) and (d) respectively show the room temperature anharmonic three-phonon scattering rates and phonon group velocity of ZrSe$_3$ monolayer, plotted as a function of frequency.

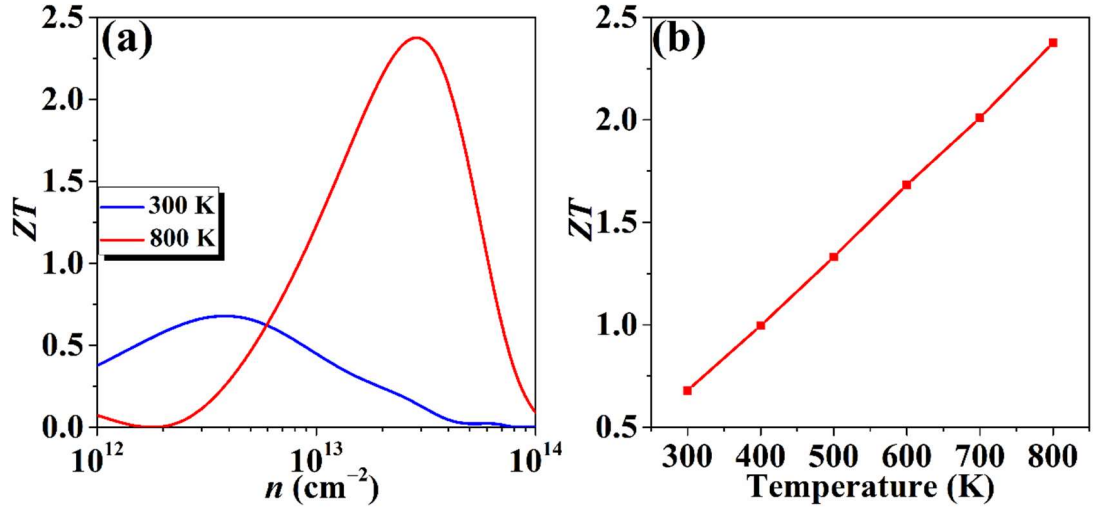

**Figure 5.** (a) *n*-type *ZT* values of ZrSe$_3$ monolayer as a function of carrier concentration along the *x*-direction at 300 and 800 K. (b) The temperature dependence of the *ZT* value.